\documentclass[12pt]{article}
\usepackage{amsfonts}
\usepackage{amsmath}
\usepackage{epsfig}
\usepackage{latexsym}
\usepackage{graphicx}
\usepackage{pdflscape}
\numberwithin{equation}{section}
\allowdisplaybreaks

\DeclareMathOperator{\Tr}{Tr}

\renewcommand{\thefootnote}{\fnsymbol{footnote}}
\newcounter{aff}
\renewcommand{\theaff}{\fnsymbol{aff}}
\newcommand{\affiliation}[1]{
\setcounter{aff}{#1} $\rule{0em}{1.2ex}^\theaff\hspace{-.4em}$}
\newcommand{\nn}{\nonumber \\}

\newcommand{\bra}{\langle}
\newcommand{\ket}{\rangle}
\def\({\Bigl(}
\def\){\Bigr)}
\def\cO{{\mathcal O}}

\def\l{\ell}

\begin{document}
\begin{titlepage}
\hfill\hfill
\begin{minipage}{1.2in}
DESY 13-010\\
TIT/HEP-626 
\end{minipage}

\bigskip\bigskip
\begin{center}
{\LARGE\bf Instanton Bound States in ABJM Theory}

\bigskip\bigskip
{\large Yasuyuki Hatsuda\footnote[1]{\tt yasuyuki.hatsuda@desy.de},
Sanefumi Moriyama\footnote[2]{\tt moriyama@math.nagoya-u.ac.jp} and
Kazumi Okuyama\footnote[3]{\tt kazumi@azusa.shinshu-u.ac.jp}
}\\
\bigskip\bigskip
\affiliation{1}
{\normalsize\it DESY Theory Group, DESY Hamburg\\ 
Notkestrasse 85, D-22603 Hamburg, Germany\\
and\\
Department of Physics, Tokyo Institute of Technology\\
Tokyo 152-8551, Japan} \bigskip\\
\affiliation{2} {\normalsize\it Kobayashi Maskawa Institute 
\& Graduate School of Mathematics, Nagoya University\\
Nagoya 464-8602, Japan} \bigskip\\
\affiliation{3} {\normalsize\it Department of Physics, Shinshu University\\
Matsumoto 390-8621, Japan} \bigskip\\
\end{center}

\begin{abstract}
The partition function of the ABJM theory receives non-perturbative
corrections due to instanton effects.
We study these non-perturbative corrections, including bound states of
worldsheet instantons and membrane instantons, in the Fermi-gas
approach.
We require that the total non-perturbative correction should be always
finite for arbitrary Chern-Simons level.
This finiteness is realized quite non-trivially because each bound
state contribution naively diverges at some levels.
The poles of each contribution should be canceled out in total.
We use this pole cancellation mechanism to find unknown bound state
corrections from known ones.
We conjecture a general expression of the bound state contribution.
Summing up all the bound state contributions, we find that the effect
of bound states is simply incorporated into the worldsheet instanton
correction by a redefinition of the chemical potential in the
Fermi-gas system.
Analytic expressions of the 3- and 4-membrane instanton corrections
are also proposed.

\end{abstract}

\end{titlepage}

\renewcommand{\thefootnote}{\arabic{footnote}}
\setcounter{footnote}{0}
\setcounter{section}{0}

\section{Introduction}
Recently, there is much progress in the 3-dimensional ${\mathcal N}=6$
supersymmetric Chern-Simons-matter theory with gauge group
$U(N)_k\times U(N)_{-k}$, known as the
Aharony-Bergman-Jafferis-Maldacena (ABJM) theory \cite{ABJM}.
Since the ABJM theory is believed to describe the low energy effective
theory on the $N$ multiple M2-branes, the study of this theory is a
significant step to understand M-theory.
It was shown in \cite{KWY, J, HHL} that the partition function and
vacuum expectation values of BPS Wilson loops in 3d supersymmetric
Chern-Simons theories including the ABJM theory are reduced to
finite-dimensional matrix integrals after the localization technique
\cite{P} is applied.
This reduction allows us to use many matrix model techniques.
In \cite{DMP1}, the famous $N^{3/2}$ degrees of freedom of M2-branes
predicted from the AdS/CFT correspondence \cite{KT} was reproduced
from the free energy of the matrix model in the leading 't Hooft
expansion.
After that, it was shown in  \cite{FHM} that the all-genus summation
results in the expression by the Airy function if the worldsheet
instanton corrections are dropped out.

In \cite{MP}, another formulation to study the ABJM matrix model was
introduced.
The crucial point is that the ABJM partition function is regarded as
the partition function of an ideal Fermi-gas system.\footnote{
Various ideas related to this formulation were proposed, including the
mirror expression in \cite{KWY,O}, generalizations to less
supersymmetric case \cite{MPinteract} or Wilson loop expectation
values \cite{KMSS}, and studies \cite{AHS} of the ABJ partition 
functions \cite{ABJ}.
}
This equivalence enables us to apply the standard methods in quantum
mechanics and statistical mechanics.
Since the Chern-Simons level $k$ plays a role of the Planck constant
$\hbar=2\pi k$, the WKB expansion corresponds to the expansion around
$k=0$, which is the strong coupling limit in the ABJM theory.
In \cite{MP}, such a WKB expansion of the partition function was
studied in detail, and further developed in \cite{CM}.
The Fermi-gas formalism is also powerful to compute the partition
function exactly.
In \cite{HMO1, PY}, the exact ABJM partition function at $k=1$ was
computed up to some values of $N$.
In \cite{HMO2}, we further developed the method to compute the
partition function and obtain its exact values for higher $N$ at
various levels $k$.
Using these exact values, we can extract the non-perturbative
corrections to the partition function by subtracting the perturbative
Airy function contribution.
The result is recapitulated in \eqref{Jnp}, with the notation
explained below.

In the Fermi-gas formalism, it is useful to consider the grand
partition function and the grand potential.
The grand potential is divided into the following three contributions,
\begin{align}
J(k, \mu)=J^{\rm (pert)}(k, \mu)+J^{\rm (np)}(k, \mu)+J^{\rm (osc)}(k, \mu),
\end{align}
where $J^{\rm (pert)}(k,\mu)$, $J^{\rm (np)}(k,\mu)$ and
$J^{\rm (osc)}(k,\mu)$ represent the perturbative, the
non-perturbative and the oscillatory parts, respectively.
The oscillatory part $J^{\rm (osc)}(k,\mu)$ arises because the
perturbative and the non-perturbative parts do not respect the
periodicity in $\mu$ that the grand partition function originally has. 
However, as discussed in \cite{HMO2}, this oscillatory part can be
removed by the deformation of the contour when we consider the
partition function.
Thus we neglect it in this paper.
The perturbative part was computed in \cite{MP} and is given by
\begin{align}
J^{\rm (pert)}(k, \mu)=\frac{C(k)}{3}\mu^3+B(k)\mu+A(k),
\end{align}
where
\begin{align}
B(k)=\frac{k}{24}+\frac{1}{3k}, \quad
C(k)=\frac{2}{\pi^2 k}. 
\end{align}
The small $k$ expansion of $A(k)$ was first computed in \cite{MP}, and
then its exact form was proposed in \cite{KEK}.
Using this result, one can easily rederive the perturbative partition
function including all the $1/N$ corrections \cite{FHM}.

In the present paper, we study the non-perturbative part in more
detail.
It is known that there are at least two kinds of instantons that
induce the non-perturbative corrections in the ABJM theory.
One is called the worldsheet instanton, and the other the membrane
instanton.
From the viewpoint of the Type IIA string theory on
$AdS_4\times\mathbb{CP}^3$, which is holographically dual to the ABJM
theory in the 't Hooft limit, the worldsheet instanton comes from the
fundamental string wrapping the holomorphic cycle
$\mathbb{CP}^1\subset\mathbb{CP}^3$ \cite{DMP1,Cagnazzo:2009zh}, while
the membrane instanton comes from the D2-brane wrapping the Lagrangian
submanifold
$\mathbb{RP}^3\subset\mathbb{CP}^3$ \cite{DMP2}.
The membrane instanton correction and the worldsheet instanton
correction to the grand potential take the following forms \cite{MP}:
\begin{align}
J^{\rm M2}(k,\mu)=\sum_{\l=1}^\infty
\Big[a_\l(k)\mu^2+b_\l(k)\mu+c_\l(k)\Big]e^{-2\l\mu},\quad
J^{\rm WS}(k,\mu)=\sum_{m=1}^\infty d_m(k)e^{-\frac{4m\mu}{k}}.
\label{eq:JWS}
\end{align}
When $k\ll N^{1/5}$, it is more appropriate to consider the M-theory
on $AdS_4\times S^7/\mathbb{Z}_k$ as the holographic dual.
From this M-theory perspective, the membrane instantons and the
worldsheet instantons are both coming from the M2-branes wrapping some
3-cycles ${\cal M}$ and ${\cal W}$ in $S^7/\mathbb{Z}_k$, where
${\cal M}$ and ${\cal W}$ descend to $\mathbb{RP}^3$ and
$\mathbb{CP}^1$ in the Type IIA picture, respectively.
As suggested in \cite{HMO2} and further explained in \cite{CM}, in
addition to such two kinds of instantons, it is natural to expect that
there is a bound state of $\ell$-membrane instanton and $m$-worldsheet
instanton which gives rise to the correction of order
$\cO(e^{-(2\l+\frac{4m}{k})\mu})$.
In the M-theory picture, such an $(\ell,m)$ bound state corresponds to
an M2-brane wrapping $\ell$ times on ${\cal M}$ and $m$ times on
${\cal W}$.
Therefore the non-perturbative part is expected to have the following
form,
\begin{align}
J^{\rm (np)}(k,\mu)
=\sum_{\substack{\l,m=0 \\ (\l,m) \ne (0,0)}}^\infty
f_{\l, m}(k,\mu) \exp\left[-\(2\l+\frac{4m}{k}\)\mu\right],
\end{align}
where the membrane instanton correction corresponds to $m=0$, while
the worldsheet instanton correction to $\l=0$.

Our goal is to determine the coefficient $f_{\l , m}(k,\mu)$ of the
$(\ell,m)$ bound state correction.
In the Fermi-gas approach, the membrane instanton corrections appear
as exponentially suppressed corrections in $\mu$, while the worldsheet
instanton corrections appear as quantum mechanical instantons.
This means that the membrane instantons can be captured in the
standard WKB analysis but the worldsheet instantons are invisible in
the small $k$ expansion.
Fortunately, since the worldsheet instanton corrections are related to
the known results of the topological string on local $\mathbb{F}_0$,
we can systematically compute $d_m(k)$, as pointed out in
\cite{HMO2}.
The bound state contributions are also invisible in the WKB analysis.
Moreover, it is unclear how these corrections are understood in the
topological string context.
There are no results on the bound state correction $f_{\l , m}(k,\mu)$
so far.

To overcome this difficulty, we employ the following strategy.
We require that the non-perturbative correction $J^{\rm (np)}(k,\mu)$
should be always finite for arbitrary value of $k$.
This fact was first observed in \cite{HMO2} for some integers $k$.
This requirement is realized very non-trivially because each bound
state contribution $f_{\l,m}(k,\mu)$ naively has poles at some values
of $k$.
As an example, let us consider the $\cO(e^{-2\mu})$ correction at
$k=2n$ ($n\in\mathbb{Z}$).
At $k=2n$, the $\cO(e^{-2\mu})$ correction comes from the two
contributions: the 1-membrane instanton and the $n$-worldsheet
instanton.
There are no bound state contributions at this order.
As pointed out in \cite{HMO2}, the $n$-worldsheet instanton
contribution $f_{0,n}(k,\mu)$ diverges at $k=2n$.
Thus the 1-membrane instanton contribution $f_{1,0}(k,\mu)$ must also
diverge at $k=2n$, and the sum of these two contributions becomes
finite.
Using this pole cancellation mechanism in addition to the matching of
a few coefficients of the WKB expansion, we proposed in \cite{HMO2}
the explicit form of the 1-membrane instanton correction
$f_{1,0}(k,\mu)$.
Recently, the 2-membrane instanton contribution $f_{2,0}(k,\mu)$ was
also proposed in the similar strategy \cite{CM}.
Here we use the pole cancellation mechanism to find the general bound
state correction $f_{\l,m}(k,\mu)$.

Remarkably, from this pole cancellation mechanism and some other input
data, we can conjecture the explicit form of $f_{\l, m}(k,\mu)$.
Our conjecture resolves all the discrepancies observed in our previous
paper \cite{HMO2}.
After summing up all the bound state contributions, we find that the
bound state contributions are incorporated into the worldsheet
instanton correction by a simple redefinition of the chemical
potential,
\begin{align}
J^{\rm (np)}(k, \mu)=J^{\rm M2}(k,\mu)+J^{\rm WS}(k,\mu_{\rm eff})~,
\label{eq:J^np}
\end{align}
where the redefined chemical potential $\mu_{\rm eff}$ is given by
\begin{align}
\mu_{\rm eff}=\mu+\frac{1}{C(k)} \sum_{\l=1}^\infty a_\l(k) e^{-2\l \mu}.
\label{eq:mu_eff}
\end{align}
Thus our remaining task is to determine the membrane instanton
correction $J^{\rm M2}(k,\mu)$ and the worldsheet instanton correction
$J^{\rm WS}(k,\mu)$.
As mentioned above, $J^{\rm WS}(k,\mu)$ can be systematically
computed up to any desired order once the weighted sum of Gopakumar-Vafa (GV) invariants on local $\mathbb{F}_0$
is known.
The membrane instanton
correction $J^{\rm M2}(k,\mu)$ will be also determined order by order.
In this paper, we further propose the 3- and 4-membrane instanton
corrections explicitly.

The organization of this paper is as follows.
In the next section, we shall first review the Fermi gas formalism.
In section 3 we shall embark on our study of the bound state
corrections.
Using the results found in section 3, in section 4 we proceed to
higher membrane instanton corrections.
Finally we conclude in section 5.
Appendices A and B are devoted to the summary of the known results of the
non-perturbative corrections of grand potential and the instanton
coefficients.
In appendix C, we derive the relation between $\mu$ and
$\mu_{\rm eff}$ for even $k$.

\section{Review of the Fermi-gas approach}
Let us start by reviewing the Fermi-gas formalism describing the ABJM
partition function proposed in \cite{MP} and the non-perturbative
corrections to the grand potential \cite{MP, HMO2, CM}.

\subsection{Fermi-gas formalism}
The partition function of ABJM theory on a round $S^3$ is written as a
matrix integral \cite{KWY}
\begin{align}
Z(k,N)=\frac{1}{(N!)^2}\int\frac{d^N\mu_i}{(2\pi)^N}\frac{d^N\nu_i}{(2\pi)^N}
\frac{\prod_{i<j}[2\sinh\frac{\mu_i-\mu_j}{2}]^2
[2\sinh\frac{\nu_i-\nu_j}{2}]^2}
{\prod_{i,j}[2\cosh\frac{\mu_i-\nu_j}{2}]^2}
\exp\biggl[\frac{ik}{4\pi}\sum_i(\mu_i^2-\nu_i^2)\biggr]~.
\label{ABJMmatrix}
\end{align}
The key observation in \cite{MP} is that \eqref{ABJMmatrix} can be
recast into the partition function of a Fermi-gas system
\begin{align}
Z(k,N)=\frac{1}{N!}\sum_{\sigma\in S_N}
(-1)^{\epsilon(\sigma)}\int\frac{d^Nq}{(2\pi )^N}
\prod_i\rho(q_i,q_{\sigma(i)}),
\end{align}
with the density matrix given by
\begin{align}
\rho(q_1,q_2)=\frac{1}{k}\frac{1}{\sqrt{2\cosh\frac{q_1}{2}}}
\frac{1}{2\cosh\frac{q_1-q_2}{2k}}
\frac{1}{\sqrt{2\cosh\frac{q_2}{2}}}.
\end{align}
After introducing the chemical potential $\mu$ or the fugacity
$z=e^\mu$, the grand partition function
\begin{align}
\Xi(k,\mu)=1+\sum_{N=1}^\infty Z(k, N)e^{\mu N},
\label{grand}
\end{align}
is written as the well-known form for the Fermi-gas system,
\begin{align}
\Xi(k,\mu)=\prod_{j=0}^\infty (1+e^{-(E_j-\mu)}),
\end{align}
where $E_j$ is the energy spectrum of the one-particle Hamiltonian defined by
\begin{align}
\hat{\rho}=e^{-\hat{H}},\quad \bra q_1| \hat{\rho} | q_2\ket =\rho(q_1,q_2).
\end{align}
Mathematically, the grand partition function is nicely expressed as a Fredholm determinant for the kernel $\rho$,
\begin{align}
\Xi(k,\mu)=\det(1+z\rho)
=\exp\left[-\sum_{n=1}^\infty\frac{(-z)^n}{n}\Tr\rho^n\right].
\end{align}
We are interested in the non-perturbative corrections to the grand
potential defined by
\begin{align}
J(k,\mu)=\log\Xi(k,\mu).
\end{align}
Note that the partition function can be reconstructed from the grand
partition function by
\begin{align}
Z(k,N)=\oint\frac{dz}{2\pi i}\frac{\Xi(k,\mu)}{z^{N+1}}
=\int_{-\pi i}^{\pi i}\frac{d\mu}{2\pi i}e^{J(k,\mu)-N\mu}.
\end{align}

\subsection{Non-perturbative corrections to the grand potential}
As mentioned in the introduction, the grand potential receives the
non-perturbative corrections.
The worldsheet instanton correction $J^{\rm WS}(k,\mu)$ can be
systematically computed by using the results of the topological string
on local $\mathbb{F}_0$,
\begin{align}
J^{\rm WS}(k, \mu)
=\sum_{g=0}^\infty \sum_{n,d=1}^\infty 
n_d^g\left(2\sin\frac{2\pi n}{k}\right)^{2g-2}
\frac{(-1)^{dn}}{n}e^{-\frac{4dn\mu}{k}}~,
\label{WSsum}
\end{align}
where $n_d^g$ denotes the weighted sum of GV
invariants on local $\mathbb{F}_0$  \cite{Katz:1999xq}
\begin{align}
n_d^g=\sum_{d_1+d_2=d}n_{d_1,d_2}^g. 
\end{align}
It is easy to see that the coefficient $d_m(k)$ is explicitly written as
\begin{align}
d_m(k)=\sum_{g=0}^\infty \sum_{d|m} n^g_d \frac{(-1)^md}{m}
\Bigl(2\sin\frac{2\pi m}{dk}\Bigr)^{2g-2},
\label{eq:d_m}
\end{align}
where $d|m$ means that $d$ is a divisor of $m$.
Using the data of GV invariants in \cite{Aganagic:2002qg, Katz:1999xq}, 
we can compute $n^g_d$ up to $(g,d)=(7,7)$.
The result is summarized in Table~\ref{tab:GV_inv}.
From these values, we can compute $d_m(k)$ up to $m=7$.

\begin{table}[tb]
\caption{The weighted sum of the GV invariants on local $\mathbb{F}_0$.}
\begin{center}
\begin{tabular}{ccccccccc}
\hline
$d$ & 1 & 2 & 3 & 4 & 5 & 6 & 7  \\
\hline
$n_d^0$ & $-4$ & $-4$ & $-12$ & $-48$ & $-240$ & $-1356$ & $-8428$   \\
$n_d^1$ & $0$ & $0$ & $0$ & $9$ & $136$ & $1616$ & $17560$   \\
$n_d^2$ & $0$ & $0$ & $0$ & $0$ & $-24$ & $-812$ & $-17340$   \\
$n_d^3$ & $0$ & $0$ & $0$ & $0$ & $0$ & $186$ & $9712$   \\
$n_d^4$ & $0$ & $0$ & $0$ & $0$ & $0$ & $-16$ & $-3156$  \\
$n_d^5$ & $0$ & $0$ & $0$ & $0$ & $0$ & $0$ & $552$   \\
$n_d^6$ & $0$ & $0$ & $0$ & $0$ & $0$ & $0$ & $-40$  \\
$n_d^7$ & $0$ & $0$ & $0$ & $0$ & $0$ & $0$ & $0$   \\
\hline
\end{tabular}
\end{center}
\label{tab:GV_inv}
\end{table}%

It is more difficult to determine the membrane instanton contribution
$f_{\l, 0}(k,\mu)=a_\l(k)\mu^2+b_\l(k)\mu+c_\l(k) $.
In \cite{MP, CM}, the WKB expansions of the coefficients $a_\l(k)$,
$b_\l(k)$ and $c_\l(k)$ were studied.
In \cite{HMO2}, the 1-membrane instanton correction $f_{1,0}(k,\mu)$
was proposed based on the requirement of the pole cancellation and the
matching of the small $k$ expansion.
Recently, the 2-membrane instanton correction $f_{2,0}(k,\mu)$ was
also proposed in a similar way \cite{CM}.
The results are summarized in \eqref{eq:1-mem}, \eqref{eq:2-mem}, and
\eqref{eq:a_n}.

Let us here sketch the pole cancellation mechanism, which is important
in our later analysis.
We first notice that the coefficients \eqref{eq:d_m}, \eqref{eq:1-mem}
and \eqref{eq:2-mem} have the poles at some values of $k$.
The crucial point is that these poles should be canceled by other
contributions including the bound states at the same order.
The final result of $J^{\rm (np)}(k,\mu)$ is finite even for these
values of $k$. 
As an example, let us consider the $\cO(e^{-2\mu})$ corrections for
even integer $k=2n$.
In this case there are two contributions: the 1-membrane instanton and
the $n$-worldsheet instanton.
At this order, there are no bound states, and the total
non-perturbative correction is given by the sum of the $1$-membrane
instanton correction and the $n$-worldsheet instanton correction.
At $k=2n$, the coefficients $b_1(k)$ and $c_1(k)$ are divergent.
One can easily check that the behavior near $k=2n$ is given by
\begin{align}
\lim_{k \to 2n}a_1(k)&=(-1)^{n-1}\frac{2}{n\pi^2},\nn
\lim_{k \to 2n}b_1(k)&=(-1)^n\frac{4}{\pi^2(k-2n)} ,\\
\lim_{k \to 2n}c_1(k)&=(-1)^{n-1}\left[-\frac{4n}{\pi^2(k-2n)^2}
-\frac{4}{\pi^2(k-2n)}+\frac{1}{3}\(\frac{1}{n}-2n\)\right],\notag
\end{align}
On the other hand, at $k=2n$, the worldsheet instanton correction
$f_{0,n}(k,\mu)$ is also divergent.
Its pole structure is given by
\begin{align}
\lim_{k \to 2n}f_{0,n}e^{-\frac{4n}{k}\mu}
=(-1)^{n-1}\left[\frac{4n}{\pi^2(k-2n)^2}+\frac{4(\mu+1)}{\pi^2(k-2n)}
+\frac{2\mu^2+2\mu+1}{n\pi^2}+w_n \right]e^{-2\mu},
\end{align}
where $w_n$ are some constants.
The poles from $f_{1,0}$ and $f_{0,n}$ precisely cancel, and we get
the final answer,
\begin{align}
\lim_{k \to 2n}(f_{1,0}e^{-2\mu}+f_{0,n}e^{-\frac{4n}{k}\mu})
=(-1)^{n-1}\left[\frac{4\mu^2+2\mu+1}{n\pi^2}+s_n\right]e^{-2\mu},
\end{align}
where $s_n=w_n+\frac{1}{3n}-\frac{2n}{3}$.

In \cite{CM}, such pole cancellation mechanism was considered at order
$\cO(e^{-4\mu})$ for odd integer $k$ to find the 2-membrane instanton
coefficients \eqref{eq:2-mem}.
For such $k$, there are no bound state contributions either, and one
can focus on the membrane instantons and the worldsheet instantons
only.
In the next section, we consider some other cases where the bound
state instanton corrections exist.
We use the pole cancellation mechanism to find the coefficients
$f_{\l,m}(k,\mu)$.

\section{Bound state corrections}
In this section, we consider the bound state contribution
$f_{\l,m}(k,\mu)$.

\subsection{$(1,n)$ bound state}
Let us first consider the $(1,n)$ bound state.
If we focus on the $\cO(e^{-4\mu})$ term for even integer $k=2n$,
there are three contributions: the 2-membrane instanton, the
$2n$-worldsheet instanton and the $(1,n)$ bound state.
Using \eqref{eq:2-mem}, the behavior of the 2-membrane instanton
correction at $k=2n$ is given by
\begin{align}
\lim_{k \to 2n} f_{2,0}e^{-4\mu}
=\left[\frac{41n}{2\pi^2(k-2n)^2}+\frac{25(2\mu+1)}{2\pi^2(k-2n)}
+\frac{-9\mu^2+4\mu}{n\pi^2}
+3\(-\frac{1}{2n}+\frac{10n}{9}\)\right]e^{-4\mu}.
\end{align}
One can check that the pole structure of the $2n$-worldsheet instanton
correction is also given by
\begin{align}
\lim_{k \to 2n} f_{0,2n}e^{-\frac{8n}{k}\mu}
=-9\left[ \frac{n}{2\pi^2(k-2n)^2}+\frac{2\mu+1}{2\pi^2(k-2n)}
+\frac{2\mu^2+\mu+1/4}{2n\pi^2}+v_n\right]e^{-4\mu},
\end{align}
where $v_n$ are some constants whose first few values are given by
\begin{align}
v_1=\frac{1}{18},\quad v_2=\frac{13}{36},\quad v_3=\frac{193}{54}.
\end{align}
Thus, the sum of these two corrections is
\begin{align}
\lim_{k \to 2n}(f_{2,0}e^{-4\mu}+f_{0,2n}e^{-\frac{8n}{k}\mu})
&=\biggl[\frac{16n}{\pi^2(k-2n)^2}+\frac{8(2\mu+1)}{\pi^2(k-2n)}
-\frac{36\mu^2+\mu+9/4}{2n\pi^2}\nn
&\hspace{0.7cm}+3\(-\frac{1}{2n}+\frac{10n}{9}\)-9v_n\biggr]
e^{-4\mu}.\label{eq:pole-1}
\end{align}
These poles must be canceled by the contribution from the $(1,n)$
bound state.

We propose the following form of the $(1,n)$ bound state coefficient:
\begin{align}
f_{1,n}(k,\mu)=-2n\pi^2a_1(k)d_n(k)~.\label{eq:f1n}
\end{align}
As we will see below, our proposal \eqref{eq:f1n} passes many
non-trivial tests.
Firstly, \eqref{eq:f1n} precisely cancels the poles in
\eqref{eq:pole-1}.
Secondly, \eqref{eq:f1n} correctly reproduces the finite part of
$J^{\rm (np)}(\mu)$ in \eqref{Jnp}.
Let us see these in more detail.
At $k=2n$, $f_{1,n}$ in \eqref{eq:f1n} behaves as
\begin{align}
\lim_{k \to 2n}f_{1,n}e^{-(2+\frac{4n}{k})\mu}
=\left[-\frac{16n}{\pi^2(k-2n)^2}-\frac{8(2\mu+1)}{\pi^2(k-2n)}
-\frac{8\mu^2}{n\pi^2}+u_n \right]e^{-4\mu},
\end{align}
where the first few values of the constants $u_n$ are given by
\begin{align}
u_1=\frac{2}{3} ,\quad u_2=-\frac{2}{3},\quad u_3=-\frac{94}{9}.
\end{align}
Thus the total result becomes finite,
\begin{align}
\lim_{k \to 2n}(f_{2,0}e^{-4\mu}
+f_{0,2n}e^{-\frac{8n}{k}\mu}+f_{1,n}e^{-(2+\frac{4n}{k})\mu})
=\left[-\frac{52\mu^2+\mu/4+9/16}{2n\pi^2}
+t_n\right]e^{-4\mu},
\end{align}
where
\begin{align}
t_n=3\(-\frac{1}{2n}+\frac{10n}{9}\)-9v_n+u_n,
\end{align}
whose first three values are given by
\begin{align}
t_1=2,\quad t_2=2, \quad t_3=-\frac{298}{9}.
\end{align}
These exactly agree with \eqref{Jnp} for $k=2,4,6$.

Moreover, at $k=4$, the $\cO(e^{-3\mu})$ terms come from $(0,3)$ and
$(1,1)$ bound states.
One finds
\begin{align}
\lim_{k \to 4}
(f_{0,3}e^{-\frac{12}{k}\mu}+f_{1,1}e^{-(2+\frac{4}{k})\mu})
=\(\frac{10}{3}+2\)e^{-3\mu}=\frac{16}{3}e^{-3\mu}.
\end{align}
Similarly, at $k=6$, the $\cO(e^{-8\mu/3})$ and $\cO(e^{-10\mu/3})$
terms come from $(0,4)+(1,1)$ and $(0,5)+(1,2)$, respectively.
One can easily check
\begin{align}
\lim_{k \to 6}
(f_{0,4}e^{-\frac{16}{k}\mu}+f_{1,1}e^{-(2+\frac{4}{k})\mu})
&=\(-8-\frac{16}{9}\)e^{-\frac{8\mu}{3}}=-\frac{88}{9}e^{-\frac{8\mu}{3}},\nn
\lim_{k \to 6}
(f_{0,5}e^{-\frac{20}{k}\mu}+f_{1,2}e^{-(2+\frac{8}{k})\mu})
&=\(\frac{244}{15}+\frac{16}{3}\)e^{-\frac{10\mu}{3}}
=\frac{108}{5}e^{-\frac{10\mu}{3}}.
\end{align} 
All these results perfectly agree with \eqref{Jnp} again.

From these checks, we strongly believe that our conjecture
\eqref{eq:f1n} is correct for any $n$.

\subsection{$(2,n)$ and $(3,n)$ bound states}
Next let us consider the $(2,n)$ bound states.
For these bound states, we have less information.
However, we here propose that $f_{2,n}(k,\mu)$ is given by the
following simple and beautiful form:
\begin{align}
f_{2,n}(k,\mu)
=\left[-2n\pi^2 a_2(k)
+\frac{1}{2!}\bigl(-2n\pi^2a_1(k)\bigr)^2\right]d_n(k).
\label{eq:f2n}
\end{align}
As a check of this proposal, let us consider the $\cO(e^{-20\mu/3})$
term at $k=3$.
This term comes from $(0,5)$ and $(2,2)$ bound states.
Thus we predict
\begin{align}
\lim_{k \to 3}
(f_{0,5}e^{-\frac{20}{k}\mu}+f_{2,2}e^{-(4+\frac{8}{k})\mu})
=\(\frac{244}{15}+\frac{16}{3}\)e^{-\frac{20\mu}{3}}
=\frac{108}{5}e^{-\frac{20\mu}{3}}.
\end{align}
Similarly, if one considers the $\cO(e^{-5\mu})$ term at $k=4$ and the
$\cO(e^{-14\mu/3})$ term at $k=6$, one obtain the predictions
\begin{align}
\lim_{k \to 4}
(f_{0,5}e^{-\frac{20}{k}\mu}+f_{1,3}e^{-(2+\frac{12}{k})\mu}+f_{2,1}e^{-(4+\frac{4}{k})\mu})
&=\(\frac{101}{5}+20+11\)e^{-5\mu}
=\frac{256}{5}e^{-5\mu},\\
\lim_{k \to 6}
(f_{0,7}e^{-\frac{28}{k}\mu}+f_{1,4}e^{-(2+\frac{16}{k})\mu}+f_{2,1}e^{-(4+\frac{4}{k})\mu})
&=\(\frac{1712}{21}+\frac{128}{3}+\frac{248}{27}\)e^{-\frac{14\mu}{3}}
=\frac{25208}{189}e^{-\frac{14\mu}{3}}.\notag
\end{align}
These correctly reproduce the result \eqref{Jnp}.
As we will see in the next section, our proposal \eqref{eq:f2n} is
also consistent with the pole structures of $b_3(k)$ and $c_3(k)$.

From the forms of $f_{1,n}$ and $f_{2,n}$, it is natural to expect
that $f_{3,n}$ takes the following form:
\begin{align}
f_{3,n}
=\left[-2n\pi^2a_3(k)
+s\bigl(-2n\pi^2a_2(k)\bigr)\bigl(-2n\pi^2a_1(k)\bigr)
+t\bigl(-2n\pi^2a_1(k)\bigr)^3\right]d_n(k),
\label{f3nst}
\end{align}
where $s$ and $t$ are some constants.
To fix $s$ and $t$, we consider the $\cO(e^{-8\mu})$ terms at $k=2$.
These terms come from $(4,0)$, $(3,1)$, $(2,2)$, $(1,3)$ and $(0,4)$.
By comparing the coefficients of $\mu^2 e^{-8\mu}$, we obtain the
constraint
\begin{align}
-\frac{2269}{\pi^2}-\frac{16}{3\pi^2}(50+27s+24t)=-\frac{2701}{\pi^2}.
\end{align}
Moreover, since the $\cO(e^{-7\mu})$ term at $k=4$ come from $(3,1)$,
$(2,3)$, $(1,5)$ and $(0,7)$, we get another condition,
\begin{align}
\frac{11882}{21}+18s+8t=\frac{4096}{7}.
\end{align}
From these conditions, $s$ and $t$ are completely fixed, and we find
\begin{align}
s=1,\quad t=\frac{1}{6}.
\end{align}
Thus \eqref{f3nst} becomes
\begin{align}
f_{3,n}=\left[ -2n\pi^2a_3(k)
+\bigl(-2n\pi^2a_2(k)\bigr)\bigl(-2n\pi^2a_1(k)\bigr)
+\frac{1}{3!}\bigl(-2n\pi^2a_1(k)\bigr)^3\right]d_n(k).
\end{align}

\subsection{Summing up all the bound states}
Let us summarize the results in the previous subsection.
For notational simplicity, we introduce
\begin{align}
A_{\l,m}\equiv-2m\pi^2a_{\l}(k).
\end{align}
In terms of $A_{\l, m}$,
our conjecture of the bound state contributions $f_{1,m}$, $f_{2,m}$
and $f_{3,m}$ in the previous subsection are written as
\begin{align}
f_{1,m}&=A_{1,m}d_m,\quad f_{2,m}=\(A_{2,m}+\frac{1}{2!}A_{1,m}^2\)d_m,\nn
f_{3,m}&=\(A_{3,m}+A_{2,m}A_{1,m}+\frac{1}{3!}A_{1,m}^3\)d_m.
\end{align}
These expressions immediately suggest us that the general coefficient
$f_{\l,m}$ is given by
\begin{align}
f_{\l,m}=\left[\sum_{(p_1,\dots,p_\l)}
\frac{1}{p_1!p_2!\cdots p_\l !}
A_{1,m}^{p_1}A_{2,m}^{p_2}\cdots A_{\l, m}^{p_\l}\right]d_m,
\quad (\l=1,2,\dots),
\label{eq:bound_conj}
\end{align}
where the sum runs over all the allowed partitions of $\l$ labeled by
$(p_1,\dots, p_\l)$ satisfying
\begin{align}
p_1+2p_2+\cdots+\l p_\l=\l,\quad
p_1\geq 0,\;\;p_2\geq 0,\;\;\cdots,\;\;p_\l\geq 0.
\label{eq:partition}
\end{align}
One non-trivial check of this conjecture is to predict the coefficient
of $\mu^2 e^{-10\mu}$ at $k=2$.
Using the explicit forms of $a_\l(k)$ ($\l=1,\dots,5)$ in
\cite{CM}\footnote{
There is a typo in $a_5(k)$ in the first version of \cite{CM}.
We need to replace $9104$ in front of $\cos(5\pi k/2)$ by $9104/5$.
We are grateful to the authors of \cite{CM} for correspondence.
} and our conjecture \eqref{eq:bound_conj}, one finds
\begin{align}
&\lim_{k \to 2}\sum_{m=0}^5f_{5-m,m}e^{-(2(5-m)+\frac{4m}{k})\mu}\nn
= &\;\(\frac{31752}{5\pi^2}+\frac{4150}{\pi^2}+\frac{5760}{\pi^2}
+\frac{6888}{\pi^2}+\frac{6216}{\pi^2}+\frac{15002}{5\pi^2}\)
\mu^2e^{-10\mu}+(\text{other terms})\nn
=&\;\frac{161824}{5\pi^2}\mu^2e^{-10\mu}+(\text{other terms}).
\end{align}
This perfectly agrees with the result in \eqref{Jnp}!
In the above computation, each bound state contributes quite
non-trivially, and the sum of all these bound states reproduces the
correct answer.
Therefore this match strongly supports our conjecture
\eqref{eq:bound_conj} (at least up to $\l=4$).

If this conjecture is correct, we can perform the sum over $\l$ in
$J^{\rm (np)}(\mu)$,
\begin{align}
J^{\rm (np)}(\mu)
&=\sum_{\substack{\l,m=0\\(\l,m)\ne(0,0)}}^\infty 
f_{\l, m}(k,\mu)\exp\left[-\(2\l+\frac{4m}{k}\)\mu\right]\nn
&=\sum_{\l=1}^\infty f_{\l,0}e^{-2\l \mu}
+\sum_{m=1}^\infty d_me^{-\frac{4m}{k}\mu}\sum_{\l=0}^\infty
e^{-2\l \mu}\sum_{(p_1,\dots,p_\l)}\frac{1}{p_1!p_2!\cdots p_\l!}
A_{1,m}^{p_1}A_{2,m}^{p_2}\cdots A_{\l, m}^{p_\l}.
\end{align}
Using the summation formula
\begin{align}
\sum_{\l=0}^\infty x^\l
\sum_{(p_1,\dots,p_\l)}\frac{1}{p_1!p_2 !\cdots p_\l!}
A_{1,m}^{p_1} A_{2,m}^{p_2} \cdots A_{\l, m}^{p_\l}
=\exp\left[\sum_{n=1}^\infty x^nA_{n,m}\right],
\end{align}
we finally obtain
\begin{align}
J^{\rm (np)}(\mu)
&=\sum_{\l=1}^\infty(a_\l(k)\mu^2+b_\l(k)\mu+c_\l(k))e^{-2\l\mu}\nn
&\hspace{0.5cm}+\sum_{m=1}^\infty d_m(k)
\exp\left[-\frac{4m}{k}
\(\mu+\frac{\pi^2 k}{2}\sum_{n=1}^\infty a_n(k)e^{-2n \mu}\)\right].
\end{align}
This result means that all the bound state contributions are
incorporated into the worldsheet instanton corrections by redefining
the chemical potential $\mu$, as in \eqref{eq:J^np} and
\eqref{eq:mu_eff}.

For even $k=2n$, in particular, we can write down the analytic
relation between $\mu$ and $\mu_{\rm eff}$,
\begin{align}
\mu_{\rm eff}=\mu+(-1)^{n-1}\,2e^{-2\mu}
\,_4F_3\( 1,1,\frac{3}{2},\frac{3}{2};2,2,2;(-1)^n\,16e^{-2\mu}\).
\label{eq:mueff-evenk}
\end{align}
A derivation of this expression is presented in appendix
\ref{sec:mu-mueff}.
Similarly, for odd $k$, from the numerical value, we conjecture that
\begin{align}
\mu_{\rm eff}=\mu+e^{-4\mu}\,_4F_3\( 1,1,\frac{3}{2},\frac{3}{2};2,2,2;-16e^{-4\mu}\).
\end{align}
These relations are very similar to the relation between the function
$\kappa$ of the 't Hooft coupling $\lambda=N/k$ and the derivative of
the genus-zero free energy, found in \cite{DMP1}.

\section{Higher membrane instanton corrections}\label{sec:3-mem}
In this section, we would like to find the higher membrane instanton
corrections $f_{\l,0}(k,\mu)=a_\l(k)\mu^2+b_\l(k)\mu+c_\l(k)$.
In \cite{CM}, the explicit forms of $a_\l(k)$ up to $\l=5$ have been
proposed.
Here we first conjecture $b_3(k)$ and $c_3(k)$,
and then we will make an observation, with which we can
further proceed to discussing $b_4(k)$ and $c_4(k)$.

\subsection{Conjecture for $b_3(k)$ and $c_3(k)$}
Let us consider the $e^{-6\mu}$ terms.
For odd $k=2n+1$, such terms should vanish.
These contributions come from $(3,0)$ and $(1,2n+1)$.
This means that $f_{3,0}$ should behave, in the limit $k\to 2n+1$, as
\begin{align}
\lim_{k \to 2n+1} f_{3,0}e^{-6\mu}
=-\lim_{k \to 2n+1} f_{1,2n+1} e^{-(2+\frac{4(2n+1)}{k})\mu}
=(-1)^n \left[\frac{2n+1}{\pi(k-2n-1)}+\frac{4\mu+1}{\pi}\right]
e^{-6\mu}.
\end{align}
This is consistent with $a_3(2n+1)=0$, and give the conditions for
$b_3(k)$ and $c_3(k)$,
\begin{align}
\lim_{k \to 2n+1}b_3(k)&=(-1)^n\frac{4}{\pi},\nn
\lim_{k \to 2n+1}c_3(k)
&=(-1)^n\left[\frac{2n+1}{\pi(k-2n-1)}+\frac{1}{\pi}\right].
\label{poleodd}
\end{align}

Next let us consider $k=2n/3$ ($n \not\in 3\mathbb{Z}$).
In this case, the $\cO(e^{-6\mu})$ terms come from $(3,0)$ and
$(0,n)$.
The pole structure of $f_{0,n}e^{-4n \mu/k}$ takes the following form,
\begin{align}
\lim_{k \to 2n/3} f_{0,n} e^{-\frac{4n}{k}\mu}
=(-1)^{n-1}\left[\frac{4n}{81\pi^2(k-\frac{2n}{3})^2}
+\frac{4(3\mu+1)}{27\pi^2(k-\frac{2n}{3})}
+\frac{2\mu^2+2\mu/3+1/9}{n\pi^2}+w_n \right],
\end{align}
where
\begin{align}
w_1=\frac{1}{3},\quad
w_2=\frac{7}{6},\quad
w_4=\frac{187}{12},\quad\cdots.
\end{align}
These poles must be canceled by the poles from $b_3(k)$ and $c_3(k)$.
Thus we require the conditions at $k=2n/3$ ($n \not\in 3\mathbb{Z}$),
\begin{align}
\lim_{k \to 2n/3} b_3(k)
&=(-1)^n\frac{4}{9\pi^2(k-\frac{2n}{3})}+\cO(1),\nn
\lim_{k \to 2n/3} c_3(k)
&=(-1)^n\left[\frac{4n}{81\pi^2(k-\frac{2n}{3})^2}
+\frac{4}{27\pi^2(k-\frac{2n}{3})}+\cO(1)\right].
\label{pole2n/3}
\end{align}

Finally let us consider $k=2n$.
For such $k$, $(3,0)$, $(2,n)$, $(1,2n)$ and $(0,3n)$ bound states
contribute to the $\cO(e^{-6\mu})$ terms.
Since we have already known the analytic expressions of $f_{2,n}$,
$f_{1,2n}$ and $f_{0,3n}$ for small $n$, we find the pole structure of
each term at $k=2n$,
\begin{align}
\lim_{k \to 2n}f_{2,n}e^{-(4+\frac{4n}{k})\mu}&=(-1)^{n-1}\biggl[
\frac{104n}{\pi^2(k-2n)^2}+\frac{4(26\mu+9)}{\pi^2(k-2n)}
+\frac{4(13\mu^2-4\mu)}{n\pi^2}+x_n\biggr]e^{-6\mu},\nn
\lim_{k \to 2n}f_{1,2n}e^{-(2+\frac{8n}{k})\mu}&=(-1)^{n-1}\biggl[
\frac{36n}{\pi^2(k-2n)^2}+\frac{18(4\mu+1)}{\pi^2(k-2n)}
+\frac{72\mu^2}{n\pi^2}+y_n\biggr]e^{-6\mu},\\
\lim_{k \to 2n}f_{0,3n}e^{-\frac{12n}{k}\mu}&=(-1)^{n-1}\biggl[
\frac{328n}{27\pi^2(k-2n)^2}+\frac{328(3\mu+1)}{27\pi^2(k-2n)}
+\frac{2(738\mu^2+246\mu+41)}{27n\pi^2}+z_n\biggr]e^{-6\mu},\notag
\end{align}
with some constants $x_n$, $y_n$ and $z_n$.
The requirement of the pole cancellation gives the conditions
\begin{align}
\lim_{k \to 2n}b_3(k)
&=(-1)^n\left[\frac{1912}{9\pi^2(k-2n)}+\cO(1)\right],\nn
\lim_{k \to 2n}c_3(k)
&=(-1)^n\left[\frac{4108n}{27\pi^2(k-2n)^2}
+\frac{1786}{27\pi^2(k-2n)}+\cO(1)\right]~.
\label{eq:c3_pole_even_n}
\end{align}
\eqref{poleodd}, \eqref{pole2n/3}  and \eqref{eq:c3_pole_even_n} are
the pole conditions that $b_3(k)$ and $c_3(k)$ should satisfy.

Here we propose the explicit forms of $b_3(k)$ and $c_3(k)$ satisfying
both the above pole structures and the correct WKB expansions in
\cite{CM},
\begin{align}
b_3(k)&=\frac{4}{\pi^2 k}
\left[13\cos\(\frac{\pi k}{2}\)+5\cos\(\frac{3\pi k}{2}\)\right]\nn
&\hspace{0.5cm}+\frac{1}{3\pi}\csc\(\frac{3\pi k}{2}\)
(241+405\cos(\pi k)+222\cos(2\pi k)+79\cos(3\pi k)+9\cos(4\pi k)),\nn
c_3(k)&=\(\frac{\pi^2}{6}+\frac{\pi^2 k^2}{48}\)a_3(k)
+\frac{\pi^4 k^2}{12} a_1(k)^3-\frac{1}{2}k\sec\(\frac{\pi k}{2}\)
-\frac{6}{\pi}\cot\(\frac{\pi k}{2}\)\cos\(\frac{\pi k}{2}\)\nn
&\hspace{0.5cm}+\frac{1}{\pi}\cot\(\frac{3\pi k}{2}\)
\left[\cos\(\frac{\pi k}{2}\)+\frac{10}{9}\cos\(\frac{3\pi k}{2}\)
+\cos\(\frac{5\pi k}{2}\)\right]\nn
&\hspace{0.5cm}+k\left[3\cos\(\frac{\pi k}{2}\)
+\frac{79}{6}\cos\(\frac{3\pi k}{2}\)
+\frac{5}{2}\cos\(\frac{5\pi k}{2}\)\right]\nn
&\hspace{0.5cm}+k\csc^2\(\frac{3\pi k}{2}\)
\left[95\cos\(\frac{\pi k}{2}\)+\frac{343}{6}\cos\(\frac{3\pi k}{2}\)
+19\cos\(\frac{5\pi k}{2}\)\right].
\label{eq:b3c3}
\end{align}
These results reproduce the coefficients of $e^{-6\mu}$ for $k=2,4$
correctly,
\begin{align}
\lim_{k \to 2} \sum_{m=0}^3 f_{3-m,m} e^{-(2(3-m)+\frac{4m}{k})\mu}
&= \biggl[\frac{736\mu^2-304\mu/3+154/9}{3\pi^2}-32\biggr]e^{-6\mu} ,\nn
\lim_{k \to 4} \sum_{m=0}^3 f_{3-m,2m} e^{-(2(3-m)+\frac{8m}{k})\mu}
&=  \left[ -\frac{736\mu^2-304\mu/3+154/9}{6\pi^2}+32 \right]e^{-6\mu}~.
\end{align}

We should comment on our proposal \eqref{eq:b3c3}.
First of all, the pole conditions and the coefficients of the WKB
expansions in \cite{CM} do not fix the forms of $b_3(k)$ and $c_3(k)$
completely.
We need to make assumptions for the forms of these functions, and then
try to fix the coefficients in the ansatz.
For $b_3(k)$, we can fix all the unknown coefficients in the ansatz
from the pole conditions and a part of the WKB data, and check that
the obtained result give the higher coefficients of the WKB expansion
correctly.
For $c_3(k)$, however, there are too many unknown parameters in the
ansatz to fix them from the pole conditions and the WKB data.
Nevertheless, we conjecture the explicit form \eqref{eq:b3c3}.
To find it, we follow the criteria that $c_3(k)$ should not have so
complicated coefficients.
This criteria seems to be natural from the results for $c_1(k)$ and
$c_2(k)$.
Moreover, in the next subsection, we will find a very non-trivial
relation (see \eqref{bcrel}).
Using this relation, $c_\l(k)$ is automatically determined by
$b_\l(k)$, and this relation is valid for $\l=3$ as well as $\l=1,2$.
However, there remains an important problem to check whether our
proposal for $c_3(k)$ indeed reproduces the higher WKB coefficients
correctly or not.

\subsection{Some observations}
Concerning the forms of membrane instanton corrections, we find some
curious observations.
As seen in the previous section, the summation over the bound states
results in the worldsheet instanton correction with the redefined
chemical potential $\mu_{\rm eff}$.
This fact suggests us to rewrite the other part of the grand potential
also in terms of $\mu_{\rm eff}$.
Namely, we would like to rewrite the sum of the perturbative part and
the membrane instanton part
\begin{align}
&J^{\rm (pert)}(k,\mu)+ J^{\rm M2}(k,\mu)\nn
&\quad=
\left[\frac{C(k)}{3}\mu^3+B(k)\mu +A(k)\right]
+\Big[\mu^2J_a(k,\mu)+\mu J_b(k,\mu)+J_c(k,\mu)\Big]~,
\label{sumpertM2}
\end{align}
where we have introduced the notation
\begin{align}
J_a(k,\mu)=\sum_{\l=1}^\infty a_\l(k)e^{-2\l \mu},\quad
J_b(k,\mu)=\sum_{\l=1}^\infty b_\l(k)e^{-2\l \mu},\quad
J_c(k,\mu)=\sum_{\l=1}^\infty c_\l(k)e^{-2\l \mu}.
\end{align}
When writing in terms of $\mu_{\rm eff}=\mu+J_a(k,\mu)/C(k)$, the
$\mu^2J_a$ term in \eqref{sumpertM2} is absorbed into the perturbative
part and the final result becomes
\begin{align}
J^{\rm (pert)}(k,\mu)+J^{\rm M2}(k,\mu)&=J^{\rm (pert)}(k,\mu_{\rm eff})+\mu_{\rm eff}
\widetilde{J}_b(k,\mu_{\rm eff})+\widetilde{J}_c(k,\mu_{\rm eff}),
\end{align}
where $\widetilde{J}_b(k,\mu_{\rm eff})$ and
$\widetilde{J}_c(k,\mu_{\rm eff})$ are given by
\begin{align}
\widetilde{J}_b(k,\mu_{\rm eff})
&=\sum_{\l=1}^\infty\widetilde{b}_\l(k)e^{-2\l\mu_{\rm eff}}
=J_b(k,\mu)-\frac{J_a(k,\mu)^2}{C(k)},\\
\widetilde{J}_c(k,\mu_{\rm eff})
&=\sum_{\l=1}^\infty\widetilde{c}_\l(k)e^{-2\l\mu_{\rm eff}}
=J_c(k,\mu)-\frac{J_a(k,\mu) J_b(k,\mu)}{C(k)}
-\frac{B(k)}{C(k)}J_a(k,\mu)+\frac{2J_a(k,\mu)^3}{3C(k)^2}.\notag
\end{align}
The coefficients $\widetilde{b}_\l(k)$ and $\widetilde{c}_\l(k)$ are
related to $a_\l(k)$, $b_\l(k)$ and $c_\l(k)$.
The first three coefficients are explicitly found to be
\begin{align}
\widetilde{b}_1(k)&=\frac{2}{\pi}\cot\(\frac{\pi k}{2}\)
\cos\(\frac{\pi k}{2}\),\nn
\widetilde{b}_2(k)&=\frac{1}{\pi}\cot(\pi k)\bigl(4+5\cos(\pi k)\bigr),\nn
\widetilde{b}_3(k)&=\frac{4}{3\pi}
\cot\(\frac{3\pi k}{2}\)
\cos\(\frac{\pi k}{2}\)
\bigl(13+19\cos(\pi k)+9\cos(2\pi k)\bigr),
\end{align}
and
\begin{align}
\widetilde{c}_1(k)
&=\frac{1}{\pi}\cot\(\frac{\pi k}{2}\)\cos\(\frac{\pi k}{2}\)
+\frac{k}{4}\csc^2\(\frac{\pi k}{2}\)\cos\(\frac{\pi k}{2}\)
\bigl(3-\cos(\pi k)\bigr),\nn
\widetilde{c}_2(k)
&=\frac{1}{4\pi}\cot(\pi k)\bigl(4+5\cos(\pi k)\bigr)
+\frac{k}{16}\csc^2(\pi k)\bigl(16+25\cos(\pi k)-5\cos(3\pi k)\bigr)
,\nn
\widetilde{c}_3(k)
&=\frac{2}{9\pi}\cot\(\frac{3\pi k}{2}\)\cos\(\frac{\pi k}{2}\)
\bigl(13+19\cos(\pi k)+9\cos(2\pi k)\bigr) \nn
&\hspace{0.5cm}+\frac{k}{12}\csc^2\(\frac{3\pi k}{2}\)
\cos\(\frac{\pi k}{2}\)
\bigl(54+87\cos(\pi k)+53\cos(2\pi k)
\nn&\hspace{5cm}
-2\cos(3\pi k)-13\cos(4\pi k)-15\cos(5\pi k)\bigr) .
\end{align}
Very surprisingly, we find that $\widetilde{c}_\l(k)$ can be expressed
by $\widetilde{b}_\l(k)$,
\begin{align}
\widetilde{c}_\l(k)=- k^2 \frac{\partial}{\partial k} 
\left(\frac{\widetilde{b}_\l(k)}{2\l k}\right),\quad (\l=1,2,3).
\label{bcrel}
\end{align}
We stress that, if we use this relation, $\widetilde c_\l(k)$ is
completely determined by $\widetilde b_\l(k)$ and subsequently the
original coefficients $b_\l(k)$ and $c_\l(k)$ can be reconstructed by
these redefined functions.
Therefore we expect that the redefined function $\widetilde b_\l(k)$
plays a more fundamental role in studying the higher membrane instanton
corrections.

Furthermore, we note that $\widetilde b_\l(k)$ has an interesting
structure.
Considering that the poles appearing in $d_m(k)$
originate from the genus-zero GV invariants $n^0_d$, we should be able
to separate the divergences in $\widetilde b_\l(k)$ and express them
in terms of $n^0_d$.
In fact we find that $\widetilde b_\l(k)$ can be written as
\begin{align}
\widetilde b_1(k)&=-\frac{1}{2\pi}\cos\Bigl(\frac{\pi k}{2}\Bigr)
\biggl[n^0_1\cot\Bigl(\frac{\pi k}{2}\Bigr)\biggr],\nn
\widetilde b_2(k)&=-\frac{1}{\pi}\cos(\pi k)
\biggl[n^0_2\cot\Bigl(\frac{\pi k}{2}\Bigr)
+\frac{n^0_1}{2^2}\cot(\pi k)\biggr],\nn
\widetilde b_3(k)&=-\frac{3}{2\pi}\cos\Bigl(\frac{3\pi k}{2}\Bigr)
\biggl[n^0_3\cot\Bigl(\frac{\pi k}{2}\Bigr)
+\frac{n^0_1}{3^2}\cot\Bigl(\frac{3\pi k}{2}\Bigr)+8\sin(\pi k)\biggr].
\end{align}
From this, we conjecture that the general structure of
$\widetilde b_\l(k)$ is given by
\begin{align}
\widetilde b_\l(k)=-\frac{\l}{2\pi}\cos\Bigl(\frac{\pi k\l}{2}\Bigr)
\biggl[\sum_{d|\l}\frac{n^0_d}{(\l/d)^2}\cot\Bigl(\frac{\pi k\l}{2d}\Bigr)
+\sum_{n=1}^{N_\l}\beta_{\l,n}\sin(\pi kn)\biggr].
\label{generaltildeb}
\end{align}
This structure of the divergent part solves the pole cancellation condition
explicitly.
In fact, using \eqref{bcrel}, we find that the divergence coming from
\begin{align}
\bigl(\mu_{\rm eff}\widetilde b_\l(k)+\widetilde c_\l(k)\bigr)
e^{-2\l\mu_{\rm eff}}
+d_m(k)e^{-4m\mu_{\rm eff}/k},
\end{align}
at $k=2m/\l$ cancels each other.

Besides, we find that the remaining finite part of 
$\widetilde{J}^{\rm (np)}(k,\mu_{\rm eff})
\equiv J(k,\mu(\mu_{\rm eff}))-J^{\rm (pert)}(k,\mu_{\rm eff})$ 
at integers $k$ is given by
\begin{align}
\widetilde{J}^{\rm (np)}(k_{\rm odd},\mu_{\rm eff})
&=\sum_{\ell=1}^\infty\biggl[
\frac{\sum_{d|\ell}(d^3n^0_d)}{16\pi^2\ell^3k}
\bigl(1+4\ell\mu_{\rm eff}+8(\ell\mu_{\rm eff})^2\bigr)
+\sum_n\frac{(-1)^nkn\beta_{2\ell,n}}{4}
\biggr](-1)^{\ell}e^{-4\ell\mu_{\rm eff}}
\nonumber\\
&\hspace{2cm}
+\sum_{m=1}^\infty h(k,m)[n^g_d]e^{-4m\mu_{\rm eff}/k},
\nonumber\\
\widetilde{J}^{\rm (np)}(k_{\rm even},\mu_{\rm eff})
&=\sum_{\ell=1}^\infty\biggl[
\frac{\sum_{d|\ell}(d^3n^0_d)}{2\pi^2\ell^3k}
\bigl(1+2\ell\mu_{\rm eff}+2(\ell\mu_{\rm eff})^2\bigr)
+\sum_n\frac{kn\beta_{\ell,n}}{4}
\biggr](-1)^{k\ell/2}e^{-2\ell\mu_{\rm eff}}
\nonumber\\
&\hspace{2cm}
+\sum_{m=1}^\infty h(k,m)[n^g_d]e^{-4m\mu_{\rm eff}/k},
\label{Jgeneral}
\end{align}
where $h(k,m)[n^g_d]$ is a linear functional of $n^g_d$ with
complicated coefficients.
Note especially that the $\beta$ dependence at even integers $k$ is a
simple sum, while that at odd integers $k$ is an alternating sum.
This means the finite parts for integers $k$ give only two conditions.

Moreover, the condition for even integers is essentially same as the condition
from the next-to-leading WKB coefficient of $\widetilde{b}_\l(k)$.
Let us see this fact in more detail.
We notice that the leading and the next-to-leading WKB coefficients of $\widetilde{b}_\l(k)$ 
are related to the $g=0$ GV invariants $n^0_d$ and the coefficient $\beta_{\l,n}$ through
\begin{align}
\widetilde{b}_\l^{(0)}=-\frac{1}{\l^2 \pi^2} \sum_{d|\l} d^3 n^0_d, \qquad
\widetilde{b}_\l^{(1)}=\sum_{d|\l} \( \frac{d}{12}+\frac{d^3}{8}\) n^0_d-\frac{\l}{2} \sum_{n} n \beta_{\l,n}, 
\label{eq:bt_WKB}
\end{align}
where the WKB expansion of $\widetilde{b}_\l(k)$ is defined by
\begin{align}
\widetilde{b}_\l(k)=\frac{1}{k} \sum_{m=0}^\infty \widetilde{b}_\l^{(m)} k^{2m}.
\end{align}
Recalling that the leading and the next-to-leading WKB coefficients of $a_\l(k)$, $b_\l(k)$ and $c_\l(k)$ 
are easily computed from the functions $J_0(\mu)$ and $J_1(\mu)$ in \cite{MP},
one can compute $\widetilde{b}_\l^{(0)}$ and $\widetilde{b}_\l^{(1)}$ for any $\l$.
This means that one can fix $n^0_d$ and $\sum_n n \beta_{\l,n}$ for any desired $d$ and $\l$ order by order by solving \eqref{eq:bt_WKB}.

We also note that the non-perturbative corrections to the grand
potential at $k=1,2$ are particularly simple because the contributions
for $g>1$ in $J^{\rm WS}$ trivially vanishes,
\begin{align}
\widetilde{J}^{\rm (np)}(1,\mu_{\rm eff})
&=\sum_{\l=1}^\infty\biggl[
\frac{\sum_{d|\l} (d^3 n^0_d)}{16\l} 
\(\frac{8(\l\mu_{\rm eff})^2+4\ell\mu_{\rm eff}+1}
{\l^2\pi^2}-\frac{1}{2}\)
+\frac{\sum_{d|\l}d\bigl(n^0_d-n_{2d}^0\bigl(1-(-1)^{\frac{\ell}{d}}\bigr)\bigr)}{16\l}\nn
&\hspace{2cm}+\frac{\sum_{d|\l}(dn^1_d)}{\l} 
+\sum_n \frac{(-1)^n n\beta_{2\l,n}}{4} 
\biggr]
(-1)^\l e^{-4\l \mu_{\rm eff}},\nn
\widetilde{J}^{\rm (np)}(2,\mu_{\rm eff})
&=\sum_{\l=1}^\infty \biggl[
\frac{\sum_{d|\l} (d^3 n^0_d)}{4\l}
\(\frac{2(\l\mu_{\rm eff})^2+2\l\mu_{\rm eff}+1}{\l^2 \pi^2}-\frac{1}{2}\)\nn
&\hspace{2cm}+ \frac{\sum_{d|\l}(dn^1_d)}{\l}
+\sum_n \frac{n \beta_{\l,n}}{2} ~\biggr](-1)^\l e^{-2\l \mu_{\rm eff}}.
\end{align}

Finally, let us study the redefinition of $\mu_{\rm eff}$ itself.
We can also express $\mu$ in terms of $\mu_{\rm eff}$:
\begin{align}
\mu=\mu_{\rm eff}+\frac{1}{C(k)} \sum_{\l=1}^\infty e_\l(k) e^{-2\l \mu_{\rm eff}}~,
\end{align}
where the coefficients $e_\l(k)$ are related to $a_\l(k)$ and fixed
order by order,
\begin{align}
e_1(k)&=\frac{4}{\pi^2 k} \cos\(\frac{\pi k}{2}\),\nn
e_2(k)&=\frac{2}{\pi^2 k} \cos(\pi k),\nn
e_3(k)&=\frac{8}{3\pi^2 k} \cos\(\frac{3\pi k}{2}\)
\bigl(2+3\cos(\pi k)\bigr) ,\\
e_4(k)&=\frac{1}{\pi^2 k} \cos(2\pi k)
\bigl(17+32\cos(\pi k)+16\cos(2\pi k)\bigr),\nn
e_5(k)&=\frac{4}{5\pi^2 k} \cos\(\frac{5\pi k}{2}\)
\bigl(101+190\cos(\pi k)+140\cos(2\pi k)
+60\cos(3\pi k)+10\cos(4\pi k)\bigr). \notag
\end{align} 
These look simpler than $a_\l(k)$.

\subsection{Further conjecture for $b_4(k)$ and $c_4(k)$}
If we believe the relation \eqref{bcrel} holds for any $\l$, the whole
system of $b_\l(k)$ and $c_\l(k)$ are determined explicitly by
$\widetilde b_\l(k)$, which consists of fewer coefficients.
To determine $\widetilde b_4(k)$,
we make an ansatz for $\widetilde b_4(k)$ similar to the form of $a_4(k)$, as
in the case for $\l=1,2,3$.
If we require that, in the sectors with 
\begin{itemize}
\item $(4,0)$ and $(0,n)$ bound states at $k=n/2$ (for $n=1,3,5,7$),
\item $(4,0)$, $(2,n)$ and $(0,2n)$ bound states
at $k=n$ (for $n=1,3$),
\item $(4,0)$, $(3,1)$, $(2,2)$, $(1,3)$ and $(0,4)$ bound states
at $k=2$,
\end{itemize}
the poles cancel and the finite terms for the integer $k$ match with
our fitting results \eqref{Jnp}, we find explicitly that
\begin{align}
\widetilde b_4(k)=\frac{1}{2\pi}\cot(2\pi k)
\bigl(164+288\cos(\pi k)+197\cos(2\pi k)
+96\cos(3\pi k)+32\cos(4\pi k)\bigr).
\end{align}
Note that as we have studied in \eqref{generaltildeb},
$\widetilde b_4(k)$ can also be put into the form
\begin{align}
\widetilde b_4(k)&=-\frac{2}{\pi}\cos(2\pi k)
\biggl[n^0_4\cot\Bigl(\frac{\pi k}{2}\Bigr)
+\frac{n^0_2}{2^2}\cot(\pi k)
+\frac{n^0_1}{4^2}\cot(2\pi k)
+48\sin(\pi k)+16\sin(2\pi k)\biggr].
\end{align}
From this, we can reconstruct $b_4(k)$ and $c_4(k)$ without
difficulty.
Since the expressions are lengthy, we shall list here only the
expansion around $k=0$,
\begin{align}
b_4(k)&=\frac{33635}{12\pi^2k}-\frac{17165k}{6}+\frac{127339\pi^2k^3}{90}
-\frac{12223\pi^4k^5}{27}+\frac{1945973\pi^6k^7}{18900}+\cO(k^9),
\nn
c_4(k)&=\Bigl(-\frac{1225}{6}+\frac{54113}{144\pi^2}\Bigr)\frac{1}{k}
+\Bigl(\frac{32821}{144}+175\pi^2\Bigr)k
+\Bigl(-\frac{57512\pi^2}{135}-\frac{217\pi^4}{3}\Bigr)k^3\nn
&\quad+\Bigl(\frac{5405089\pi^4}{22680}+\frac{175\pi^6}{9}\Bigr)k^5
+\Bigl(-\frac{5057377\pi^6}{64800}-\frac{1343\pi^8}{360}\Bigr)k^7
+\cO(k^9).
\label{eq:c4}
\end{align}
The first two coefficients are found to match the results of \cite{MP}
correctly.
However, these matches are automatic from the above cancellation
mechanism.\footnote{We thank the authors of \cite{CM} for checking
that the higher coefficients in \eqref{eq:c4} also perfectly agree
with the ones obtained from the WKB expansion in \cite{CM}.}

We can repeat the same analysis for $b_5(k)$ and $c_5(k)$.
However, in an ansatz of $\widetilde b_5(k)$ similar to $a_5(k)$,
there are more coefficients than conditions and we cannot determine
them explicitly.
One way to see it is that, as we noted below \eqref{Jgeneral}, the
finite values of $J(k,\mu)$ for integers $k$ gives two conditions to
determine two coefficients $\beta_{4,n}$ in $\widetilde b_4(k)$, while
$\widetilde b_5(k)$ is expected to have more coefficients which cannot
be determined only by one condition in \eqref{Jnp}.

\section{Summary and discussions}
In this paper, we have studied the non-perturbative corrections to the
ABJM partition function in the Fermi-gas approach.
These non-perturbative corrections contain the contributions from the
bound states of the worldsheet instantons and the membrane instantons,
which originate from the M2-branes wrapping two different cycles
${\cal M}$ and ${\cal W}$ on the gravity dual side.

We have proposed an expression for the coefficients of the bound
states $f_{\ell,m}(k,\mu)$ and the 3- and 4-membrane instanton
corrections explicitly.
We have found that the summation over the bound states is beautifully
incorporated into the worldsheet instanton correction by redefining
the chemical potential $\mu$.
We believe that our results on the bound states open up a
new era in understanding instanton effects in M-theory since, unlike the
worldsheet instanton or the membrane instanton, the bound states were
not accessible from any established methods such as the topological
strings or the WKB analysis in the Fermi-gas.
Our method is basically the same as our previous work \cite{HMO2} by
requiring the singularities coming from each instanton contribution to
cancel and matching the finite results and the WKB expansions.
We have also found that the non-perturbative correction to the grand
potential can be reexpressed in a simpler form by using the redefined
chemical potential.
This result implies a deep hidden structure in the ABJM partition function.

We shall conclude our paper with discussions on the further
directions.

Firstly, we have only used a limited number of data to propose
analytic expressions, which are considered to have infinite coefficients potentially.
It would be important to check our results in more
details by increasing the data such as the WKB expansions and the
finite results in the grand potential.

Secondly, after determining the coefficients up to 4-membrane
instanton, we would like to head for a more general structure of the
membrane instanton correction.
For example, we have found that the redefinition of the chemical
potential simplifies the expression substantially and the new
expression enjoys an interesting relation among the coefficients.
It would be interesting to understand the physical meaning of the
redefinition.
Perhaps, $\mu_{\rm eff}$ might be interpreted as a certain
generalization of the flat coordinate on the quantum moduli space of
local $\mathbb{F}_0$, including the membrane instanton corrections.
The similarity pointed out below \eqref{eq:mueff-evenk} may be useful
to clarify the interpretation of $\mu_{\rm eff}$.
Along this line, we would like to raise a question whether $\mu$ or
$\mu_{\rm eff}$ is more ``fundamental''.
As we have seen, most of our general arguments are expressed more
simply in $\mu_{\rm eff}$.
However, though we observed \cite{HMO2} in \eqref{Jnp} a similarity in
coefficients between $J(1,\mu)$ and $J(2,\mu)$ and between $J(3,\mu)$
and $J(6,\mu)$ even for terms not multiplied by $1/\pi^2$, this
similarity is not found when expressed in $\mu_{\rm eff}$.

Also, our general expression \eqref{generaltildeb} for 
$\widetilde b_\l(k)$, guided by the pole cancellation mechanism,
contains the coefficients $\beta_{\ell,n}$ and the relation between
$\mu$ and $\mu_{\rm eff}$ contains the coefficients in $e_\l(k)$ or
$a_\l(k)$.
It is interesting to observe that these coefficients share some kind
of integral property.
It would be important to understand whether they can be expressed in
terms of the GV or other invariants.
We would like to study higher membrane instanton effects to clarify
the structure.

Finally, it would be also interesting to study the relation, if any, between 
the worldsheet instanton corrections and the membrane instanton
corrections under the inversion of the coupling $k/2\rightarrow 2/k$
along the lines of \cite{Lockhart:2012vp}.

\vskip5mm
\centerline{\bf Acknowledgements}
\vskip3mm
\noindent
We are grateful to Shinji Hirano, Kazuo Hosomichi, Hiroaki Kanno,
Soo-Jong Rey, Kazuhiro Sakai, Masaki Shigemori, Seiji Terashima, and
especially Flavio Calvo and Marcos Marino, for useful discussions.
S.M. is grateful to Yukawa institute for theoretical physics for
hospitality, where part of this work was done.
The work of Y.H. is supported in part by the JSPS Research Fellowship
for Young Scientists, while the work of K.O. is supported in part by
JSPS Grant-in-Aid for Young Scientists (B) \#23740178.

\appendix
\section{$J^{\rm (np)}(k, \mu)$ at $k=1,2,3,4,6$}
Here we summarize the non-perturbative corrections to the grand
potential at $k=1,2,3,4,6$.
These are the updated data, part of which are previously listed in
\cite{HMO2}.
These coefficients can be fixed by the numerical fitting from the
exact values of the partition function.
See \cite{HMO2} for more detail:
\begin{align}
J^{\rm (np)}(1,\mu)&=\biggl[\frac{4\mu^2+\mu+1/4}{\pi^2}\biggr]e^{-4\mu}
+\biggl[-\frac{52\mu^2+\mu/2+9/16}{2\pi^2}+2\biggr]e^{-8\mu}
\nonumber\\
&\quad+\biggl[\frac{736\mu^2-152\mu/3+77/18}{3\pi^2}-32\biggr]e^{-12\mu}\nn
&\quad+\biggl[-\frac{2701\mu^2-13949\mu/48+11291/768}{\pi^2}+466\biggr]e^{-16\mu} \nn
&\quad+\biggl[\frac{161824\mu^2-317122\mu/15+285253/300}{5\pi^2}-6720\biggr]e^{-20\mu} \nn
&\quad+\biggl[-\frac{1227440\mu^2-2686522\mu/15+631257/80}{3\pi^2}-\frac{292064}{3}\biggr]e^{-24\mu} 
+\cO(e^{-28\mu}), \nonumber\\
J^{\rm (np)}(2,\mu)&=\biggl[\frac{4\mu^2+2\mu+1}{\pi^2}\biggr]e^{-2\mu}
+\biggl[-\frac{52\mu^2+\mu+9/4}{2\pi^2}+2\biggr]e^{-4\mu}
\nonumber\\
&\quad+\biggl[\frac{736\mu^2-304\mu/3+154/9}{3\pi^2}-32\biggr]e^{-6\mu} \nn
&\quad+\biggl[-\frac{2701\mu^2-13949\mu/24+11291/192}{\pi^2}+466\biggr]e^{-8\mu}\nn
&\quad+\biggl[\frac{161824\mu^2-634244\mu/15+285253/75}{5\pi^2}-6720\biggr]e^{-10\mu} \nn
&\quad+\biggl[-\frac{1227440\mu^2-5373044\mu/15+631257/20}{3\pi^2}-\frac{292064}{3}\biggr]e^{-12\mu} 
+\cO(e^{-14\mu}), \nonumber\\
J^{\rm (np)}(3,\mu)&=\frac{4}{3}e^{-\frac{4}{3}\mu}
-2e^{-\frac{8}{3}\mu}
+\biggl[\frac{4\mu^2+\mu+1/4}{3\pi^2}+\frac{20}{9}\biggr]e^{-4\mu}
-\frac{88}{9}e^{-\frac{16}{3}\mu}+\frac{108}{5}e^{-\frac{20}{3}\mu}\nn
&\quad+\left[ -\frac{52\mu^2+\mu/2+9/16}{6\pi^2}-\frac{298}{9}\right]e^{-8\mu}+\frac{25208}{189}e^{-\frac{28}{3}\mu}+\cO(e^{-\frac{32}{3}\mu}), \nonumber\\
J^{\rm (np)}(4,\mu)&=e^{-\mu}+\biggl[
-\frac{4\mu^2+2\mu+1}{2\pi^2}\biggr]e^{-2\mu}
+\frac{16}{3}e^{-3\mu}+\left[-\frac{52\mu^2+\mu+9/4}{4\pi^2}+2\right]e^{-4\mu} \nn
&\quad+\frac{256}{5}e^{-5\mu}+ \left[ -\frac{736\mu^2-304\mu/3+154/9}{6\pi^2}
+32 \right]e^{-6\mu}+\frac{4096}{7}e^{-7\mu}+\cO(e^{-8\mu}), \nonumber\\
J^{\rm (np)}(6,\mu)&=\frac{4}{3}e^{-\frac{2}{3}\mu}
-2e^{-\frac{4}{3}\mu}
+\biggl[\frac{4\mu^2+2\mu+1}{3\pi^2}+\frac{20}{9}\biggr]e^{-2\mu}
-\frac{88}{9}e^{-\frac{8}{3}\mu}+\frac{108}{5}e^{-\frac{10}{3}\mu}\nn
&\quad+\left[ -\frac{52\mu^2+\mu+9/4}{6\pi^2}-\frac{298}{9}\right]e^{-4\mu}+\frac{25208}{189}e^{-\frac{14}{3}\mu}+\cO(e^{-\frac{16}{3}\mu}).
\label{Jnp}
\end{align}

\section{Known results of membrane instanton corrections}

Here we summarize various known results of the membrane instanton coefficients.
The coefficients of 1-membrane instanton are given by
\begin{align}
a_1(k)&=-\frac{4}{\pi^2 k} \cos \Bigl(\frac{\pi k}{2} \Bigr), \nn
b_1(k)&=\frac{2}{\pi} 
\csc\Bigl(\frac{\pi k}{2}\Bigr)\cos^2\Bigl(\frac{\pi k}{2}\Bigr),
\nn
c_1(k)&=\biggl[-\frac{2}{3k}+\frac{5k}{12}
+\frac{1}{\pi}\cot\Bigl(\frac{\pi k}{2}\Bigr)
+\frac{k}{2}\csc^2\Bigl(\frac{\pi k}{2}\Bigr)\biggr]
\cos\Bigl(\frac{\pi k}{2}\Bigr),\label{eq:1-mem}
\end{align}
while those of 2-membrane instanton are
\begin{align}
a_2(k)&=-\frac{2}{\pi^2 k}\bigl(4+5\cos(\pi k)\bigr),\nn
b_2(k)&=\frac{4}{\pi^2 k}\bigl(1+\cos(\pi k)\bigr)
+\frac{1}{\pi}\csc(\pi k)\bigl(2+3 \cos(\pi k)\bigr)^2,\nn
c_2(k)&=\Bigl(-\frac{1}{3k}+\frac{1}{4\pi}\cot(\pi k)\Bigr)
\bigl(4+5\cos(\pi k)\bigr)+\frac{7k}{24}\bigl(7\cos(\pi k)-4\bigr)\nn
&\qquad+\frac{k}{4}\csc^2(\pi k)\bigl(20+21 \cos(\pi k)\bigr).
\label{eq:2-mem}
\end{align}
The coefficients $a_\ell(k)$ are determined up to $\ell=5$ as
\begin{align}
a_3(k)&=-\frac{8}{3\pi^2k}\cos\Bigl(\frac{\pi k}{2}\Bigr)
\bigl(19+28\cos(\pi k)+3\cos(2\pi k)\bigr),\notag\\
a_4(k)&=-\frac{1}{\pi^2k}
\bigl(364+560\cos(\pi k)+245\cos(2\pi k)+48\cos(3\pi k)+8\cos(4\pi k)\bigr),
\nn
a_5(k)&=-\frac{8}{5\pi^2k}\cos\Bigl(\frac{\pi k}{2}\Bigr)
\bigl(2113+3374\cos(\pi k)+1751\cos(2\pi k)+525\cos(3\pi k)\nn
&\qquad+145\cos(4\pi k)+25\cos(5\pi k)+5\cos(6\pi k)\bigr).
\label{eq:a_n}
\end{align}

\section{Relation of $\mu$ and $\mu_{\rm eff}$ for even $k$}\label{sec:mu-mueff}
In this appendix, we derive \eqref{eq:mueff-evenk} for even integer $k=2n$.
Our goal here is to exactly perform the sum
\begin{align}
\sum_{\l =1}^\infty a_\l(2n) e^{-2\l \mu} .
\end{align}
Since $k a_\l(k)$ is a periodic function of $k$, the values of $a_\l(2n)$ are related to the leading WKB coefficient of $a_\l(k)$:
\begin{align}
a_\l(2n)=(-1)^{\l n} \frac{a_{\l}^{(0)}}{2n},
\end{align}
where the WKB expansion is given by
\begin{align}
a_\l(k)=\frac{1}{k}\sum_{m=0}^\infty a_\l^{(m)} k^{2m}.
\end{align}
We find that $a_{\l}^{(0)}$ is generically given by
\begin{align}
a_{\l}^{(0)}=-\frac{4^\l}{\l \pi^2} \left[ \frac{(2\l-1)!!}{\l!} \right]^2.
\end{align}
This result can be confirmed by using the large $\mu$ expansion of $J_0(\mu)$ in \cite{MP},
\begin{align}
J_0(\mu)&=\frac{e^{\mu}}{4} \,_3F_2\( \frac{1}{2},\frac{1}{2},\frac{1}{2};1,\frac{3}{2};\frac{e^{2\mu}}{16} \) 
-\frac{e^{2\mu}}{8\pi^2} \,_4F_3\( 1,1,1,1;\frac{3}{2},\frac{3}{2},2;\frac{e^{2\mu}}{16} \) \nn
&=\frac{2\mu^3}{3\pi^2}+\frac{\mu}{3}+\frac{2\zeta(3)}{\pi^2}+\sum_{\l=1}^\infty (a_{\l}^{(0)} \mu^2+b_{\l}^{(0)} \mu+c_{\l}^{(0)})e^{-2\l \mu}.
\end{align}
Therefore we finally obtain
\begin{align}
\sum_{\l =1}^\infty a_\l(2n) e^{-2\l \mu} = (-1)^{n-1} \frac{2}{n\pi^2} e^{-2\mu} \,{}_4F_3\(1,1,\frac{3}{2},\frac{3}{2};2,2,2;(-1)^n\,16 e^{-2\mu}\).
\end{align}

\end{document}